\documentclass{tcibook}
\usepackage{fancyhea}
\usepackage{work}
\usepackage{bm}       
\usepackage{graphicx}
\usepackage{hyperref}      
\usepackage{chngcntr}
\counterwithout{figure}{chapter}

\newcommand{\nc}{\newcommand}  

\def\beq{\begin{equation}}
\def\eeq#1{\label{#1}\end{equation}}
\def\eeqn{\end{equation}}

\newenvironment{Eqnarray}%
   {\arraycolsep 0.14em\begin{eqnarray}}{\end{eqnarray}}
\def\beqa{\begin{Eqnarray}}
\def\eeqa#1{\label{#1}\end{Eqnarray}}
\def\eeqan{\end{Eqnarray}}

\nc{\ra}{\rightarrow}  
\nc{\slsh}{\slash\hspace*{-0.22cm}}
\def\Re{{\cal R \mskip-4mu \lower.1ex \hbox{\it e}\,}}
\def\Im{{\cal I \mskip-5mu \lower.1ex \hbox{\it m}\,}}

\nc{\vev}[1]{ \left\langle {#1} \right\rangle }
\nc{\bra}[1]{ \langle {#1} | }
\nc{\ket}[1]{ | {#1} \rangle }
\nc{\fb}{\,{\rm fb}^{-1}}
\nc{\ev}{{\rm eV}}
\nc{\kev}{{\rm keV}}
\nc{\Mev}{{\rm MeV}}
\nc{\gev}{{\rm GeV}}
\nc{\tev}{{\rm TeV}}
\nc{\mev}{{\rm MeV}}

\def\del{\partial}
\def\Dslash{\not{\hbox{\kern-4pt $D$}}}
\def\dslash{\not{\hbox{\kern-2pt $\del$}}}
\def\pslash{\not{\hbox{\kern-2pt $p$}}}
\def\ETmiss{ \not{\hbox{\kern-4pt $E$}}_T }

\def\msb{{\bar{\ssstyle M \kern -1pt S}}}

\begin{document}

\def\bibname{References}

\bibliographystyle{utphys}  

\raggedbottom

\pagenumbering{roman}

\parindent=0pt
\parskip=8pt
\setlength{\evensidemargin}{0pt}
\setlength{\oddsidemargin}{0pt}
\setlength{\marginparsep}{0.0in}
\setlength{\marginparwidth}{0.0in}
\marginparpush=0pt


\pagenumbering{arabic}

\renewcommand{\chapname}{chap:cc}
\renewcommand{\chapterdir}{.}
\renewcommand{\arraystretch}{1.25}
\addtolength{\arraycolsep}{-3pt}
\newcommand\aj{Astron. J.}
\newcommand\nat{Nature}
\newcommand\apjl{Astrophys. J.}
\newcommand\pasa{PASA}


\renewcommand*\thesection{\arabic{section}}

\newcommand{\be}{\begin{equation}}
\newcommand{\bs}{\begin{split}}
\newcommand{\bea}{\begin{eqnarray}}
\newcommand{\eea}{\end{eqnarray}}
\newcommand{\jcap}{JCAP}
\newcommand{\aap}{Astron. \& Astrophys.}
\newcommand{\prd}{Phys. Rev. D}
\newcommand{\apjs}{Astrophys. J. Supp.}
\newcommand{\apj}{Astrophys. J.}
\newcommand{\mnras}{Mon. Not. Roy. Ast. Soc.}
\newcommand{\om}{\Omega_m}
\newcommand{\ode}{\Omega_{de}}
\newcommand{\Oe}{\Omega_e}
\newcommand{\onu}{\omega_\nu}
\newcommand{\ocd}{\omega_c}
\newcommand{\ob}{\omega_b}

\chapter*{Exploiting Cross Correlations and Joint Analyses}

\begin{center}\begin{boldmath}



\begin{center}

\begin{large} {\bf Convener: J.~Rhodes}\footnote{jason.d.rhodes@jpl.nasa.gov; Jet Propulsion Laboratory, California Institute of Technology}
 \end{large}

S.~Allen,
B.A.~Benson,
T.~Chang,
R.~de~Putter,
S.~Dodelson,
O.~Dor\'{e},
K.~Honscheid,
E.~Linder,
B.~A.~Mazin,
B.~M\'{e}nard,
J.~Newman,
B.~Nord,
E.~Rozo,
E.~Rykoff,
A.~Vallinotto,
D.~Weinberg

\end{center}



\end{boldmath}\end{center}


\section{Overview}
\label{sec:cc-overview}
The nature of the dark energy thought to be causing the accelerating expansion of the Universe is one of the most compelling questions in all of science.  Any of the explanations for the accelerated expansion, whether a new field, a negative pressure fluid, or a modification to General Relativity will signal new physics and have a profound effect on our understanding of the Universe.  The current observational constraints on dark energy and modifications to gravity still allow for a large range of models and theoretical explanations. Given the importance of dark energy, we must attack the problem from a variety of angles, taking advantage of \textbf{cross-correlations} between and \textbf{joint analyses} of different probes, missions, wavelengths, and surveys, to enable the most stringent cosmological constraints.

Dark energy has two observational consequences. The first is an accelerated expansion history as encoded in the redshift-distance relationship.
This is indeed the way that dark energy was discovered in the 1990s.  Of the primary observational probes of dark energy, Type Ia Supernovae and the baryonic acoustic oscillations (BAO) signal in large scale structure are best suited to measure the expansion history of the Universe~\cite{distance}.  The second effect of dark energy is that the growth of large-scale structure is inhibited as the attractive force of gravity works against the repulsive nature of dark energy.  Weak gravitational lensing, galaxy clusters, and redshift space distortions (RSD) are all well-suited for measuring the growth of structure~\cite{growth}.


The theme of this paper is that combining these probes allows for more accurate (less contaminated by systematics) and more precise (statistically since there is information in the cross-correlations) measurements of dark energy. We  explore ways in which disparate surveys can significantly enhance our knowledge of dark energy and our ability to control systematics and other errors. We also present arguments for inter-agency cooperation to fully empower the community to make use of the data that will be available in the coming decades.

We present a mix of work in progress and suggestions for future scientific efforts.  The paper is organized as follows.  In \S\ref{sec:xcorr}, we describe the general framework of cross-correlation measurements and the basic information provided by each probe of dark energy.  We then (\S\ref{sec:cc-probes}) describe the ways in which the multiple dark energy probes provided by imaging-based (also referred to as ``photometric'') dark energy experiments are correlated, and how these correlations may be profitably exploited. Then, in \S\ref{sec:spec}, we explore new opportunities for cross-correlation studies that are possible when information from spectroscopic and imaging experiments are combined. \S\ref{sec:mul} broadens the horizon to discuss observations at other wavelengths that will help extend the reach of the optical surveys.
A brief discussion of the benefits of joint analysis of data is provided in \S\ref{sec:cc-joint}.

\section{Cross-Correlations}

\label{sec:xcorr}

The canonical attack on dark energy features four independent probes: Type Ia supernovae, weak gravitational lensing, the BAO/RSD signatures in the observed Large-Scale Structure, and galaxy clusters \cite{growth,distance}. If the probes were uncorrelated with one another, each could be studied independently with its own set of systematics; the results of the probes could be combined (after checking for consistency) by multiplying the individual likelihoods together. Inconsistency in the probes would signal a defect in the model used to fit the data (assuming systematics had been properly accounted for), whereas consistency and the ensuing joint likelihood would lead to much tighter constraints on the dark energy parameters than could be delivered by a single method.  In general, each technique will yield its strongest constraints on different degeneracy directions in parameter space, so that a large range of models will be ruled out by one measurement or another, yielding tight limits in combination.  An example of the power of combining results in this way can be seen in Figure~\ref{fig:fgas}, discussed below.

Treating all likelihoods from each method as independent is appropriate only if the probes are not correlated with one another. The reality is different, largely because {\bf all the major probes are affected by the distribution of matter in the regions studied}.  This effect is weakest for Type Ia supernovae, whose apparent brightness will be affected by gravitational lensing by matter along the line of sight at a modest level.  For other probes, this dependence is more fundamental. Each probe  either relies on using galaxies as a direct tracer of mass in the universe (for BAO, large-scale-structure power spectrum measurements, and galaxy clusters), or else measures the gravitational impact of matter and uses that to constrain cosmological models (for weak lensing and RSD).

As a result, the values of cosmological observables obtained via different methods from the same region of sky will exhibit covariance; for instance, in an area where the density of galaxies is high at some redshift, the distortion of background objects by gravitational lensing will also be greater.  In general, we can measure the degree of covariance between a cosmological observable at one location and a second observable at a different location, as a function of the separation between those positions: these {\bf cross-correlation} functions can provide information not given by each observable on its own.

This language originates in studies of large-scale structure, where measurements of the two-point cross-correlation function (the excess probability of finding an object in one class at a given separation from an object in a second class, compared to random distributions) have been used to study the relationship of different populations of galaxies to the underlying dark matter distribution.  On sufficiently large scales, the two-point correlation statistics (the power spectrum or correlation function) used to measure clustering should be related to the corresponding statistic for the underlying matter distribution via a constant  ``bias'' parameter that describes the relative overdensities of galaxies compared to dark matter; cross-correlations between probes can be critical for determining the strength of the bias.

Cross-correlation measurements can be used to recover information that each probe misses on its own.  For instance, understanding the nature of the large-scale structure bias is key for many cosmological applications of large-scale structure (particularly power-spectrum measurements and RSD -- the impact on BAO is minimal), since theories make clean predictions only for the statistics of the matter distribution, but we observe the clustering of visible objects.  Weak gravitational lensing measurements can help by constraining weighted integrals of the clustering of dark matter.  The combination of information on the observed clustering of galaxies, the integrated weak lensing signal, and the cross-correlation between the two can break the degeneracy between bias and dark matter clustering in the linear regime to provide strong constraints on the growth of structure.

In turn, while the strength of weak lensing depends on the clustering of dark matter, the observed signal is difficult to interpret without information on the redshifts of the sources of the lensing effect as well as of background objects.  Cross-correlations between the lensing signal and the positions of  galaxies whose redshifts are measured in the course of large-scale-structure studies can break degeneracies and allow more powerful constraints than from lensing alone.

The use of galaxy clusters as a probe of dark energy benefits from lensing measurements -- which can enable direct measurement of average masses of clusters selected based on other properties -- as well as large-scale-structure measurements, as determination of the bias of clusters by cross-correlation with other populations can provide another constraint on mass with different weighting \cite{majumdar}.  Lensing magnification effects on Type Ia Supernovae are subtle, but knowledge of which lines of sight should have more or less lensing allows them to be measured more readily, mitigating their impact on supernova cosmology studies.

In the sections following, we consider a few of the most important applications of cross-correlations and synergies between multiple probes in more detail.

\section{Photometric Experiments}
\label{sec:cc-probes}

Large photometric (imaging-based) experiments are a key ingredient in the assault on dark energy \cite{facilities}. Data from these projects will be used to provide multiple, complementary constraints on dark energy.  They will also be used to select targets for spectroscopic experiments.  As described above, the combination of likelihood information from multiple probes can yield stronger constraints than any single method on its own; cross-correlations will increase the constraining power of these experiments further.  In this section, we will first quickly summarize the information provided by these experiments, and then focus on two examples of how analyses that incorporate cross-probe information and methods can be used to constrain dark energy: by measuring cross-correlations between probes directly, as in galaxy-galaxy lensing, or by using the observables needed for the standard probes in new ways, as with lensing magnification.

\subsection{Imaging-based probes of dark energy}

As exemplified surveys such as the  Sloan Digital Sky Survey (SDSS), wide, pan-chromatic photometric surveys can enable a wide variety of probes of cosmology \cite{distance,growth}.   For instance, the apparent brightnesses of Type Ia supernovae depends on their distance from us, which is determined by the parameters of the cosmological model; these supernovae are found by finding objects that newly appear in repeated images of an area of sky.  The strength of weak gravitational lensing -- the apparent distortion of the shapes of background objects due to the aggregate effect of mass in the foreground -- depends on the distances to the lensing and background objects as well as the growth rate of overdensities in the universe, as that determines the amount of mass doing the lensing.  These distortions may only be measured via imaging.

The apparent abundance of clusters of galaxies -- the most massive virialized objects in the Universe -- depends weakly on distance, but strongly on the growth rate of overdensities, as such objects are rare if matter accretes slowly; images allow clusters to be identified and their masses (via weak lensing) and redshifts to be estimated.  Finally, the observed clustering of the general populations of galaxies can provide information on both the growth rate (which affects the intensity of clustering) and neutrino masses (as sufficiently massive neutrinos will distort the observed pattern of clustering); while the BAO feature in the observed clustering pattern enables measurements of distance as a function of redshift via the ``standard ruler'' method.

It is desirable to measure both the expansion history of the Universe via distance measurements, as well as the growth rate of overdensities via one of a variety of the methods discussed, as the combination allows a test of whether the acceleration of the Universe's expansion may be due to deviations from General Relativity.  In general, both the new-generation (``Stage III'') Dark Energy Survey (DES) and Subaru HyperSuprimeCam (HSC) surveys beginning in 2013, as well as the next-generation (``Stage IV'') Large Synoptic Survey Telescope (LSST) and planned space-based experiments like Euclid and The Wide Field Infrared Survey Telescope using the $2.4m$ Astrophysics Focused Telescope Asset (WFIRST-AFTA) will employ as many of these probes as possible, since a single telescope and instrument can be used to apply all of them simultaneously.

\subsection{Exploiting Correlations Between Probes}

One example of a cross-correlation analysis made possible by imaging dark energy experiments is to take advantage of the very tight connection between gravitational lensing and large scale structure~\cite{2009ApJ...695..652B}. Consider a map of the galaxy density, with the fractional over- or under-density as a function of position on the sky denoted as $\delta^g(\vec\theta)$. We can combine this  with information from maps of the distortion of galaxy shapes, which are measured for weak lensing analyses. The distorted shapes carry information about the convergence $\kappa(\vec\theta)$, where
\begin{equation}
\kappa(\vec\theta) = \int_0^{z_s} dz \, W\left(D_A(z)\right) \delta\left(D_A(z)\vec\theta; z\right).
\label{eq:kappa}
\end{equation}
Here the integral is out to the redshift of the {\it source} galaxies, those whose shapes are used to construct the $\kappa$ map; $W(z)$ is a window function that depends on the redshift-distance relationship; $\delta$ is the over-density of matter (as opposed to $\delta^g$, which is a biased tracer of the matter field); and $D_A(z)$ is the angular diameter distance out to redshift $z$. A map of galaxy shapes determined from an imaging experiment can be transformed into a $\kappa$ map, which then encodes information about both the evolution of structure ($\delta(z)$) and the all-important redshift-distance relation, $D_A(z)$.

What are the advantages of combining the lensing map with the galaxy over-density map? Consider first the situation if the $\delta^g$ map is restricted to nearby galaxies (say in a {\it foreground} redshift bin, e.g., $0.3<z<0.5$) and the $\kappa$ map is generated from distant galaxies (say in a {\it background} bin, e.g., $0.8<z<1.0$); then
\begin{equation}
\langle \delta^g_{\rm fg} \kappa_{\rm bg} \rangle = \xi_{\rm gal-gal} \ne 0. 
\end{equation}
Far from being independent, the two maps are highly correlated, since it is precisely the over-densities associated with the foreground galaxies that distort the shapes of the background galaxies. This phenomenon was first detected over fifteen years ago~\cite{Brainerd:1995da} and is generally referred to as galaxy-galaxy lensing. The cross-correlation between the two maps, $\xi_{\rm gal-gal}$, contains information on the mass of the foreground lenses and more generally about the distribution of matter in the foreground redshift bin.  The resulting information on the clustering bias of foreground populations is precious: throwing it away by only treating the lensing information separately from the galaxy position information would waste valuable evidence about cosmic structure and yield weaker constraints on dark energy models.

The lensing map is correlated with maps of galaxy densities in more ways than this. Cross-correlating the observed background galaxy density with either the foreground or background shear map leads to
\begin{equation}
\langle \delta_{\rm bg} \kappa \rangle = \xi_{\rm mag};
\end{equation}
again the two fields are correlated, this time because $\kappa$ affects the flux arriving from the background galaxies.  This effect of {\it magnification} is rapidly emerging as a powerful tool, complementing the more traditional shape measurements; the next subsection highlights some of the recent developments and the potential of magnification.


The improved constraints on cosmological models obtained from combining lensing and large-scale-structure measurements were assessed by  \cite{Hearin:2012}, who find that lensing analyses yield $\sim 20\%$ smaller errors on dark energy parameters when all of the cross-correlations between lensing convergence and large-scale-structure measurements are included.  In particular,  they include  measurements of the power spectrum of galaxies for a series of redshift bins; measurements of the power spectrum of the lensing convergence for those redshift bins; and the cross-power between the two.  Essentially, incorporating cross-correlations into analyses has the same effect on errors as covering a 50\% larger sky area, but can be done at no additional cost in survey time.

Lensing and large scale structure are not the only dark energy probes that are correlated with one another. For instance, weak lensing measurements provide a key method for obtaining mass estimates for galaxy clusters. The use of clusters to constrain the growth of structure relies on the fact that the abundance of very massive clusters is exponentially sensitive to the amplitude of matter density fluctuations at a given epoch (often parameterized by $\sigma_8$, the RMS of the density fluctuations on a scale of $8 h^{-1}$ Mpc where $h$ indicates the Hubble constant in units of 100 km s$^-1$ Mpc$^-1$). To exploit this sensitivity, we need to count the number of clusters above a given mass threshold or as a function of mass; without accurate calibration of the mass scale, this observable cannot be connected to  predictions from theory. Estimates of masses from weak lensing have therefore proven key to the use of clusters for dark energy analyses. Beyond this, though, the convergence ($\kappa$) maps generated in a lensing survey are necessarily correlated with observed cluster abundances, as the mass in the clusters is a major source of the deflections measured in lensing.
Thus, treating results from the two probes as statistically independent is incorrect. Under certain circumstances the correlations can actually increase the constraining power of imaging experiments~\cite{Takada:2007fq,Shapiro:2007cy}.

 The development of algorithms that combine information from cluster samples, large-scale-structure measurements, and weak gravitational lensing that yield as much information as possible about dark energy while simultaneously minimizing the effects of systematics is currently an active area of research~\cite{Oguri,vandenBosch:2012nq,Yoo:2012vm}. As an example, \cite{Yoo:2012vm} advocates measuring the large scale clustering of a set of galaxies in a given redshift bin, and to then use galaxy-galaxy lensing on small scales to infer the masses of the halos hosting these galaxies.  The resulting mass estimates can be used to predict the large-scale-structure bias factor, which may then be used to map the distribution of galaxies into the distribution of dark matter (which is what theoretical models and simulations generally predict).  The resulting mass map can be used to infer the amplitude of dark matter inhomogeneities at large scales, generally described by the parameter $\sigma_8$. Carrying out this program in a variety of redshift bins would lead to measures of $\sigma_8$ as a function of redshift; i.e., a measurement of the growth of large-scale-structure to help distinguish between dark energy and modified gravity. Many ideas for combining these probes in various combinations are beginning to be implemented for Stage III surveys such as DES and HSC.

The fourth major probe of dark energy, supernovae, also  exhibit correlations with the others. Supernovae may be lensed by the intervening matter inhomogeneities, so their apparent brightnesses -- the property used as a distance indicator -- exhibit greater variance than would be expected from the the intrinsic variation in supernova properties. Several groups~\cite{Dodelson:2005zt,Quartin:2013moa} have proposed utilizing this excess ``noise'' as a tool to infer the clustering amplitude,  yielding a measurement of $\sigma_8$.

The overwhelming message from all of this work is that the four canonical dark energy probes are highly correlated with one another; these correlations must be accounted for to obtain accurate constraints on dark energy. In many cases, the correlations can in fact serve to augment the information carried by each of the probes individually, and yield stronger constraints and increased robustness to systematic effects as a result.

\subsection{Magnification}

Synergies between cosmological probes go beyond the use of direct cross-correlations.  In the case of lensing magnification -- the increase in size and flux of sources behind over-dense regions -- the same types of quantities that are measured for standard dark energy probes, namely the apparent clustering on the sky of galaxies and measurements of their shapes, can be employed to constrain dark energy in new ways.


Galaxies that happen to lie behind over-dense regions appear to us as brighter and larger than they otherwise would. The fundamental problem in detecting this effect is that their unlensed sizes and fluxes are not known {\it a priori}. If all galaxies were the same size or brightness, or even if they were drawn from a Gaussian distribution with finite width, detecting magnification would be straightforward. In reality, the distributions are closer to power laws; the distribution of brightnesses of galaxies magnified by a high-density region will follow an {\it identical} power law, and galaxies that would otherwise be too faint/small would now be included in observed samples.
 More generally, the effect of lensing magnification on brightnesses will be a shift in the average magnitudes (or log fluxes) of objects from one location to another,
\be
\delta m_{\rm obs} = C_s \delta m ,
\end{equation}
where $\delta m_{\rm obs}$ is the observed effect on the population of interest while $\delta m$ is the magnitude offset that individual objects undergo.
The proportionality constant $C_s$ ranges from $C_s=0$ for a power-law distribution (in which case the overall brightening of the detected sources is exactly compensated by the inclusion of new, fainter sources) to $C_s=1$ for a delta-function distribution in which case brightening information can be extracted on an object basis. For more details, see \cite{2010MNRAS.405.1025M}.

Since $C_S$ is  small for typical galaxy samples, the majority of magnification studies have focused on number density changes rather than changes in brightness or size. If a given patch of the sky is affected by a magnification fluctuation, the apparent density of a sample of objects in that patch will be changed, due to objects that are normally fainter than the flux limit of the sample but are brightened above it and due to the fact that the angular area covered by the objects is changed. The first robust detection of this effect around galaxies was obtained by \cite{2005ApJ...633..589S} with the SDSS, with signal measurable at up to Mpc scales.  This signal is detectable with the same statistics used for large-scale-structure studies; it essentially manifests as a cross-correlation between foreground galaxies and background objects, as well as increasing correlations in the apparent density of the background objects.

%
Recently, various studies have demonstrated that magnification can also be detected by measuring correlated changes in luminosity and size. \cite{2010MNRAS.405.1025M} used quasars (which are useless for shape-based weak lensing studies, as the emission region is unresolved) and measured the brightness change induced by foreground galaxies (see Figure~\ref{fig:magfig}.   \cite{2009PhRvL.103e1301S,2012ApJ...744L..22S} selected a sample of galaxies designed to allow extraction of a magnification signal from both their brightness and size distributions, while  \cite{2011arXiv1111.1070H} introduced a new method to estimate mean brightness changes with respect to a ``photometric fundamental plane'' relation for red sequence galaxies, exploiting the same sorts of magnitude and size information which must be measured for weak lensing studies. These papers demonstrate that a careful selection of background sources can greatly improve the sensitivity of magnification estimators; essentially, the goal is to select objects for which $C_S\rightarrow1$.


In contrast to shape-based weak lensing measurements, correlation-based magnification studies are not limited to resolved objects, and as a result can in principle make use of \emph{all} sources observed in a given survey, not just those large enough to measure shapes for.  This has its largest impact for faint objects at higher redshifts, which tend to be small in size.
Since magnification (especially those dealing only with fluxes and positions but not size) and shape measurements are affected by different types of systematics, using both of these estimators allows us to test and validate the robustness of weak lensing-based inferences.
The full potential of magnification measurements is only beginning to be explored.
%
\begin{figure}
\begin{center}
\includegraphics[width=0.8\columnwidth]{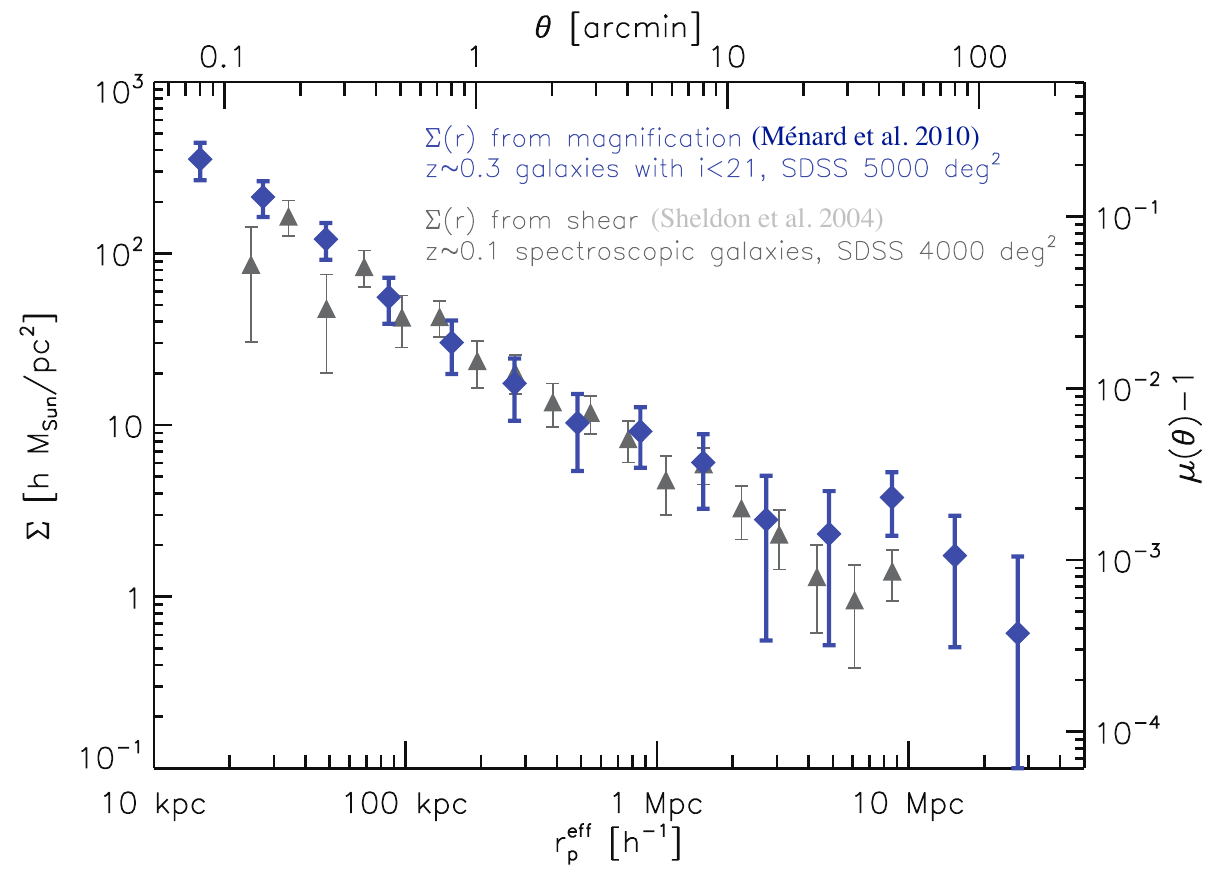}
\caption{The mean surface density of galaxies (with $i<21$) measured through the magnification of background quasars \cite{2010MNRAS.405.1025M} compared to the mean surface density of a sample of $\sim L^\star$ galaxies at $z\sim 0.1$ obtained from the gravitational shear of background galaxies from~\cite{2004AJ....127.2544S}}
  \label{fig:magfig}
  \end{center}
  \end{figure}

\section{Spectroscopic Experiments}
\label{sec:spec}

Upcoming spectroscopic surveys will not only enable exciting new probes of dark energy, but also help extend the reach of the large photometric surveys they complement. After first outlining the principal dark energy constraints coming from these surveys, we then highlight two ways in which they empower their photometric cousins.

\subsection{BAO and RSD}

Upcoming spectroscopic surveys, such as the Dark Energy Spectroscopic Instrument (DESI),  the Subaru Prime Focus Spectrograph (PFS), Euclid, and WFIRST-AFTA are optimized to extend the Baryon Acoustic Oscillation (BAO) probe to 3 dimensions and allow for a new large scale structure probe, that of Redshift Space Distortions (RSD). The velocities of distant galaxies are due not only to the Hubble expansion but also to the local gravitational forces. Therefore, when redshift is used as a distance indicator, the true position of a galaxy is mis-estimated (e.g., a galaxy on our side of a large potential well will be traveling faster away from us than the Hubble flow, and therefore we will assign it a distance farther than it actually is). These mis-estimates lead to quantifiable distortions in the two-point statistics of large scale structure. Careful study of the distortions sheds light on the cosmic velocity field which, via the continuity equation, is related to the growth of perturbations. Therefore RSD are yet another tool that can be used to infer the growth of structure. Related papers contain projections for how well BAO and RSD will constrain dark energy~\cite{distance} and neutrino properties~\cite{neutrino}.

\subsection{Cross-correlations between Redshift-Space Distortions and Lensing}

Redshift space distortions on their own are projected to be powerful probes of the growth of structure, but there is the possibility that cross-correlating them with lensing information (from photometric surveys) will lead to even more powerful probes. A number of studies~\cite{2009PhRvL.102b1302S,2011JCAP...11..039S,2011MNRAS.416.3009B,2012MNRAS.422.2904G,2012MNRAS.422.1045C,2013arXiv1307.8062K,dePutter2013nha} have been carried out as to how effective this cross-correlation will be, but these have not yet converged, so the full effect is yet to be determined. Some studies find that overlapping surveys in the same area of sky leads to substantial gains over independent surveys of the same size, but other studies find only modest gains from sky overlap, largely because the predicted gain simply from combining the complementary information on the Universe provided by each probe is so great.

Here we mention just one possibility that has already been implemented with data from the Sloan Digital Sky Survey.  Modified gravity models typically yield a modified Poisson equation relating the gravitational potential $\Phi$ to matter over-density $\delta$:
\be
\nabla^2\Phi = -4\pi G_{\rm eff} \bar\rho\delta
\label{eq:eg}
\end{equation}
where $G_{\rm eff}$ is equal to Newton's constant in dark energy models based on general relativity, but deviates from $G_N$ otherwise. These deviations are model-dependent but generically can depend on space and time. Lensing probes the gravitational potential (the left hand side of Eq.~(\ref{eq:eg})), while redshift space distortions probe the over-densities on the right-hand side. Therefore, by combining the two sets of observations, the value of $G_{\rm eff}$ can be determined~\cite{Zhang:2007nk}. In 2010, a team of astronomers carried the first such measurement using lensing data and spectroscopic data from the SDSS~\cite{Reyes:2010tr}. Upcoming surveys will greatly increase the constraining power of this test.

\subsection{Photometric Redshift Calibration}
\label{sec:cc-photoz}

One of the key areas of synergy between photometric and spectroscopic dark energy experiments will be in the area of photometric redshift calibration. Redshift information is critical for dark energy experiments, as they all rely on determining the evolution of some quantity as a function of cosmic time (for which redshift is a proxy). Distances, the growth rate of dark matter fluctuations, and the expansion rate of the Universe are all functions of redshift that can be readily calculated given a cosmological model; dark energy experiments then constrain cosmological parameters by constraining one or more of these functions.

Being able to infer redshift information from imaging alone is vital. In order not to be limited by either Poisson or ``sample'' (a.k.a. ``cosmic'') variance, the excess variance above Poisson due to the large-scale clustering of matter, most dark energy experiments study very large numbers of objects (e.g. $> 4\times10^9$ galaxies for the LSST)  over as large an area of sky as possible. It is currently infeasible to obtain redshifts via spectroscopy for so large a number of galaxies, so widely distributed; the highest-multiplexing spectrographs planned can survey $\sim5000$ objects at a time over a limited field of view, and require long exposure times to obtain redshifts for faint objects.

As a result, in the last 15 years, there have been considerable advances in determining photometric redshifts; i.e., estimates of the redshift (or the probability distribution of possible redshifts) for an object based only on imaging information, rather than spectroscopy~\cite{1985PhDT........17S,1999ASPC..191....3K}. Effectively, multi-band imaging provides a very low-resolution spectrum of an object, which can be used to determine the redshift, $z$.  A wide variety of algorithms have been developed, broadly divided into those that depend upon having a statistically complete training set of exemplar objects (exploiting advances in the field of machine learning) and those that rely on fitting models of template galaxy spectra to the observed data (see, e.g., \cite{2009ApJ...695..747B}).

A separate white paper \cite{redshift} presents estimates of the size and nature of spectroscopic samples required for designing and training photometric redshift algorithms for future experiments such as LSST.  If sufficiently complete spectroscopy could be obtained, this would serve the goal of calibration, but if there is no major change from past experience, new techniques (such as those described below) will be necessary.

\textbf{Problems with the standard methods}

Uncertainties in photo-z's are much greater than for spectroscopic redshifts; the best photometric redshifts (using $>10$ filter passbands) have attained RMS errors $\sigma_z \approx 0.007(1 + z)$ for the very brightest, highest signal-to-noise objects, while in more typical cases $0.05(1 + z)$ errors may be attained via 5-6 bands of deep imaging. Of greater concern, non-Gaussian, catastrophic (e.g. $\delta z = (|z_{phot} - z_{true}|)> 0.1(1 + z))$ photometric redshift errors occur with frequencies from 1 to 10\% or more. These errors can either occur because of problems in photometry (e.g., galaxies may overlap with each other on the sky, yielding contaminated colors for each) or due to degeneracies in photometric redshift determination (e.g., many galaxy spectra exhibit jumps at both the `Lyman break' at $912$\AA and the `Balmer break' at $4000$\AA; if one break is confused for the other, a wildly inaccurate photo-z results).

Because of these difficulties, dark energy experiments are unlikely to ever treat individual photometric redshifts as known with precision. Instead, we divide objects into redshift bins based upon their photo-z (e.g., \cite{DETF}). However, in order to obtain precision measurements of the properties of dark energy, the \emph{true redshift distribution} of the objects in each bin must be known with very high accuracy. For LSST, it is estimated that the mean redshift in each bin must be known to $\sim 2\times10^{-3}(1+z)$ for dark energy constraints not to be substantially degraded~\cite{2006ApJ...644..663Z,2006JCAP...08..008Z,2006ApJ...652..857K,2006AIPC..870...44T}. 
The true width of each redshift bin must also be known, though with somewhat less precision ($\delta \sigma_z <3 \times 10^{-3}(1 + z)$ for LSST, where $\sigma_z$ is the Gaussian sigma of the true redshift distribution). 


These targets will be extremely difficult to meet with standard photometric redshift calibration techniques. Both training-set and template-based methods ultimately depend on mapping out the relationship between galaxy colors and redshift by using some set of galaxies with both spectroscopic redshifts and photometric measurements as calibrators. However, even Stage III dark energy probes include objects too faint to obtain spectroscopic redshifts for en masse; for instance, DES \cite{2004AAS...205.6916A} extends to objects $4\times$ fainter than those for which redshifts may be obtained with $\sim70$\% success rates at the Keck Telescopes, the largest optical telescopes in the world useful for conventional multi-object spectroscopy, in one hour of observing time. Stage IV projects such as LSST, Euclid, and the Wide Field Infrared Survey-Astrophysics Focused Telescope Asset (WFIRST-AFTA) (\cite{DETF}) may reach roughly $10\times$ fainter than this limit; at least one hundred hours observation time on existing facilities would be required to obtain spectroscopic redshifts for such galaxies. Even a telescope with a 30m diameter mirror (such as the proposed TMT) would require $>10$ hours to obtain redshifts to the LSST or WFIRST-AFTA limit; furthermore, TMT would have a relatively small field of view, and hence take an incredible amount of time to cover the sky area of even a Stage III probe.

A greater difficulty is posed by the systematic failure of faint galaxy surveys to obtain redshifts for a substantial fraction of their targets. The DEEP2 Galaxy Redshift Survey, VIMOS-VLT Deep Survey, and zCOSMOS survey, for instance, obtained secure ($>95$\% confidence) redshift measurements for 40\% - 75\% of galaxy targets \cite{2005A&A...439..845L}, \cite{2009ApJS..184..218L}, \cite{2012arXiv1203.3192N} at depths of $i\sim22.5-23$; DES will utilize objects $\le 3\times$ fainter than this for cosmology measurements, and LSST will go $\sim10\times$ fainter.  If more stringent, $>99$\% confidence redshifts are required to limit systematics (as found by \cite{2012MNRAS.423..909C}), success rates for these surveys drop to $\sim20$\% - $60$\%.  This would not be an issue if the galaxies that fail to yield redshifts are a random subset, but instead failure rates depend on galaxy properties (\cite{2012arXiv1203.3192N}).

\textbf{Cross-correlation techniques}

In the past few years~\cite{2008ApJ...684...88N}, a new technique for calibrating photometric redshifts has emerged. The key idea is to use the fact that galaxies tend to cluster around one another. 
For typical populations, the probability of finding one galaxy near another galaxy is twice the expected value if there were no clustering for pairs separated by $\sim4h^{-1}$ Mpc, but only 5\% higher than random for pairs separated by $\sim20h^{-1}$ Mpc. Various populations of galaxies all cluster together both because individual dark matter halos tend to host multiple galaxies, and because the dark matter halos themselves cluster together.

Because galaxies cluster together over only relatively small distances, any observed clustering between a photometric sample and galaxies at some known redshift $z_s$ arises from galaxies in the photometric sample that have redshifts near $z_s$. Therefore, by measuring the angular cross-correlation function (the excess number of objects in one class near an object of another class on the sky, as a function of separation) between a photometric sample and a spectroscopic sample as a function of the known spectroscopic $z_s$, we can recover information about the redshift distribution of the photometric sample (hereafter denoted by $p(z)$).

As described in \cite{2008ApJ...684...88N}, these methods can determine the redshift distribution for \textbf{any} photometric sample using spectroscopy of any set of galaxies (or any other tracer of large-scale structure, e.g. quasar absorption systems), so long as spectroscopic samples span the redshift range of the photometric objects. Even if spectroscopic surveys succeed only in measuring redshifts for relatively bright objects at a given $z$, and even then in a biased or incomplete fashion - the most likely scenario given past experience - this new technique can still measure the bias and uncertainty in photometric redshifts with the accuracy required for next-generation experiments like WFIRST-AFTA and LSST with feasible amounts of spectroscopy.

If we measure only this cross-correlation, the redshift distribution will be degenerate with the strength of the intrinsic, three-dimensional clustering between the two samples. However, the two-point autocorrelation functions (the Fourier transform of the power spectrum commonly used in studies of the cosmic microwave background) of the photometric and spectroscopic samples - some of the most basic measurements that are generally made from galaxy surveys - provide sufficient information to break that degeneracy. Further care must be taken to account for magnification bias, which can be a contaminant in cross correlation photo-z tests\cite{redshift}.


Recently, Refs.~\cite{2013arXiv1303.4722M,2013MNRAS.431.3307S}  have applied cross-correlation methods to both simulations and real data with considerable success, demonstrating that redshift distributions may be recovered for photometrically-selected samples using very different tracers of large-scale structure with consistent results. Rahman et al. (in prep) are now using clustering-based redshift distribution estimates based on these methods to assess photometric redshift algorithms applied to SDSS data. The results, illustrated in Figure \ref{fig:cluster-z}, are quite encouraging. These cross-correlation-based methods are able to reveal all the intrinsic degeneracies affecting photometric redshift estimation. By revealing regions of parameter space that possess significant populations of catastrophic outliers, this new technique can allow us to use only the remaining clean regions for dark energy studies.  The work of Ref.~\cite{2013arXiv1302.0857M} has demonstrated that the constraining power of these methods may be improved further - by up to a factor of ten - by utilizing optimal estimators in power spectrum space.


A variety of estimates \cite{2008ApJ...684...88N}, \cite{2010ApJ...721..456M}, \cite{2013arXiv1302.0857M},\cite{dePutter2013nha} have found that cross-correlation techniques can yield calibration with sufficient accuracy for LSST dark energy science given a sample of ~100,000 objects with spectroscopic redshifts, spanning an area of ~100 square degrees, assuming that the redshift coverage of the spectroscopic sample is sufficient.  The combined sample from the SDSS, BOSS, and DESI surveys would far exceed these requirements over the full redshift range relevant for LSST (eBOSS should provide a sufficient sample for LSST calibration, which should be available much sooner than DESI, but with less margin for error than the latter project).  Even though these surveys have focused on the Northern sky, they will have more than sufficient overlap with the LSST area. A number of authors have suggested incorporating additional observables (e.g., information from gravitational lensing, cf. \cite{2010ApJ...720.1090Z}) that can help to constrain intrinsic galaxy clustering; this should only strengthen the resulting constraints.

\begin{figure}[thbp]
\includegraphics[width=.8\columnwidth]{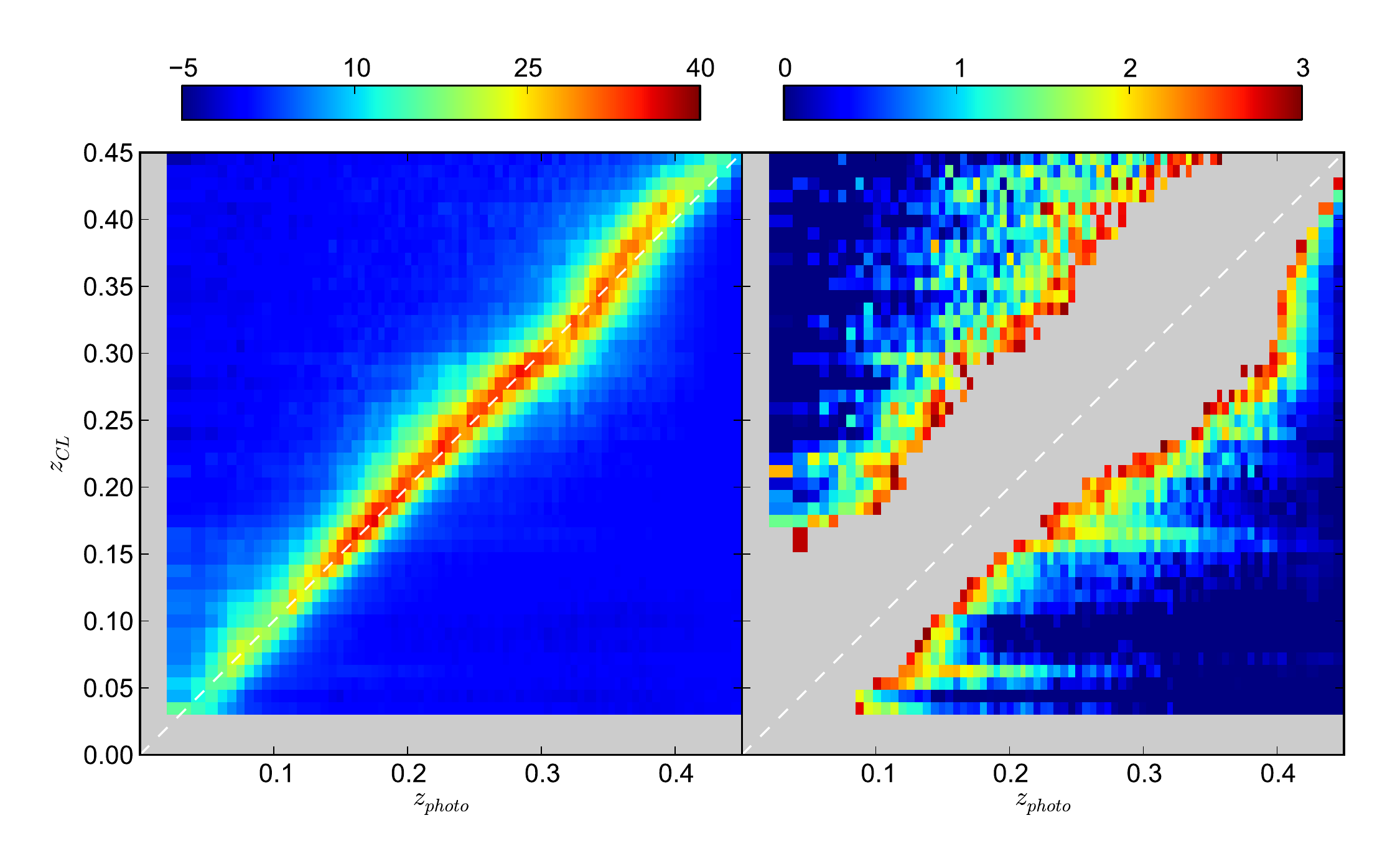}
\caption{Clustering-based reconstructed redshift distributions as a function of
best-fit photometric redshift for SDSS galaxies selected with $r<21$.
The left panel shows the full range of ${\rm d}N/{\rm d}z$ values and
shows a good agreement between the two estimators. The right panel shows the same data with a different color scheme to highlight the outliers,
revealing various features
due to intrinsic degeneracies of photometric redshifts. The horizontal
features show the effect of emission lines located close to the edge
of the response curve of various filters. Adapted from Rahman, M\'enard et al.
(in prep).}
\label{fig:cluster-z}
\end{figure}

\section{Data from Other Wavelengths}
\label{sec:mul}

Dark energy was discovered in optical surveys, and these continue to play the leading role in determining its properties, but observations at other wavelengths are becoming increasingly important as we strive for greater precision and control over systematics. Observing in the near infrared (rather than the  optical) is enabled by putting a telescope in space and has a number of advantages over optical observations alone \cite{euclidrb,2013arXiv1305.5422S}. Two other key examples are cosmic microwave background (CMB) experiments and X-ray surveys. While the CMB is most sensitive to physics at the last scattering surface at redshift $z_*=1100$, secondary anisotropies due to interactions and lensing along the line of sight have now been detected at very high significance. The importance of these secondary anisotropies for dark energy studies is only beginning to be explored, but early indications are tantalizing. X-ray observations of clusters have a longer history, and have proven important in constraining properties of galaxy clusters; measurements of the Sunyaev-Zel'dovich effect via CMB experiments provides further constraints on cluster properties.

\subsection{Cross-Correlations with CMB Experiments}
\label{sec:cc-cmb}

Maps of the cosmic microwave background temperature and polarization will exhibit significant cross-correlations with the observed density of galaxies and the lensing signals from optical surveys for multiple reasons; each provides additional information that can be used to constrain cosmology.  First, once dark energy becomes significant, CMB photons passing through foreground matter overdensities will gain energy, as the potential well they fall into becomes less shallow by the time they leave; the resulting increase of CMB temperature behind overdensities is referred to as the Integrated Sachs-Wolfe (or ISW) Effect.  This effect will cause CMB temperature to exhibit cross-correlations with maps of foreground galaxy density; this effect has now been detected for a variety of foreground galaxy samples (cf.  \cite{planckisw} and references therein), and provides an independent line of evidence for the existence of dark energy.  A second source of correlations is the Sunyaev-Zel'dovich (SZ) effect, the result of the inverse Compton scattering in energy of CMB photons that pass through galaxy clusters; it will be discussed further in the next subsection.

The most promising source of cross-correlations is gravitational lensing of the cosmic microwave background.  One of the most exciting recent discoveries in cosmology was the first 4-$\sigma$ detection~\cite{Das:2011ak} by the Atacama Cosmology Telescope of the lensing of the cosmic microwave background (CMB), followed soon thereafter by a 7-$\sigma$ detection~\cite{vanEngelen:2012va} by the South Pole Telescope and most recently a 26-$\sigma$ detection in all-sky data from the Planck satellite~\cite{Ade:2013tyw}. CMB maps provide estimates of $\kappa$ as defined in Eq.~(\ref{eq:kappa}), with the source redshift equal to the redshift of last scattering $z_*=1100$. Most of the lensing signal comes from mass at redshifts $1-5$, offering an unprecedented view of structure in the Universe at redshifts beyond those currently probed by galaxy surveys.

The CMB lensing signal should exhibit cross-correlations with both foreground galaxy density maps (as they trace the matter causing the lensing) and with weak lensing and magnification maps that use objects at intermediate redshifts as sources (as some of the mass lensing the CMB will also be between us and those source galaxies).  In fact, the weakest CMB lensing signals have been detected by cross-correlations with foreground galaxy populations \cite{sptlensing}; see Figure~\ref{fig:sptlensing}.  These cross-correlations are only beginning to be explored; the full synergies between CMB $\kappa$ measurements and planned dark energy experiments are not yet known. Applications to dark energy studies include the overlap of DES and the South Pole Telescope (SPT), or HSC and ACTPol as another example, to aid in determining the clustering bias of galaxies and the power of CMB lensing to constrain early dark energy models (described more fully in \S\ref{sec:synergy}).

The Dark Energy Survey will survey regions of the sky optically that have already been surveyed in the microwave by SPT. A simple example of how the cross-correlation of the CMB lensing map with the lensing maps from DES will aid in constraining cosmological parameters is depicted in Fig.~\ref{fig:alberto}~\cite{Vallinotto:2011ge,Vallinotto:2013eva}. The dotted lines show that DES alone will be able to constrain the cosmological parameters fairly tightly; adding in the CMB data though will significantly improve on these constraints.

\begin{figure}[thbp]
\begin{center}
\includegraphics[width=.6\columnwidth]{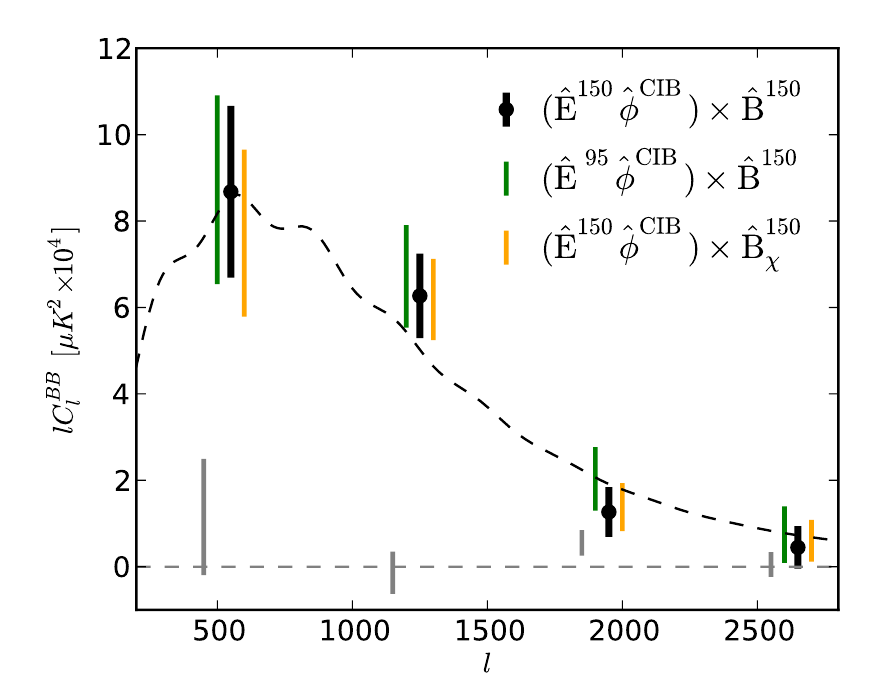}
\caption{The recent detection of the weak B-mode (curl-like) polarization signal from CMB lensing \cite{sptlensing}, made possible via a cross-correlation-based method.  The green, orange, and black error bars present the observed cross-correlation between a predicted-polarization map based on a combination of the E-mode (gradient-like) polarization and the B-mode of the polarization, with color code representing the method used for measurement, in bins of multipole number $l$.  The dashed line indicates the expected signal for the fiducial dark energy model adopted; the grey error bars show the result of a successful null test of the method.  Even though it is $\sim 100\times$ weaker than the E-mode lensing effect previously detected, the B-mode lensing effect is detected here at 7.7$\sigma$ significance by taking advantage of cross-correlation techniques. }
\label{fig:sptlensing}
\end{center}
\end{figure}

\begin{figure}[thbp]
\begin{center}
\includegraphics[width=.6\columnwidth]{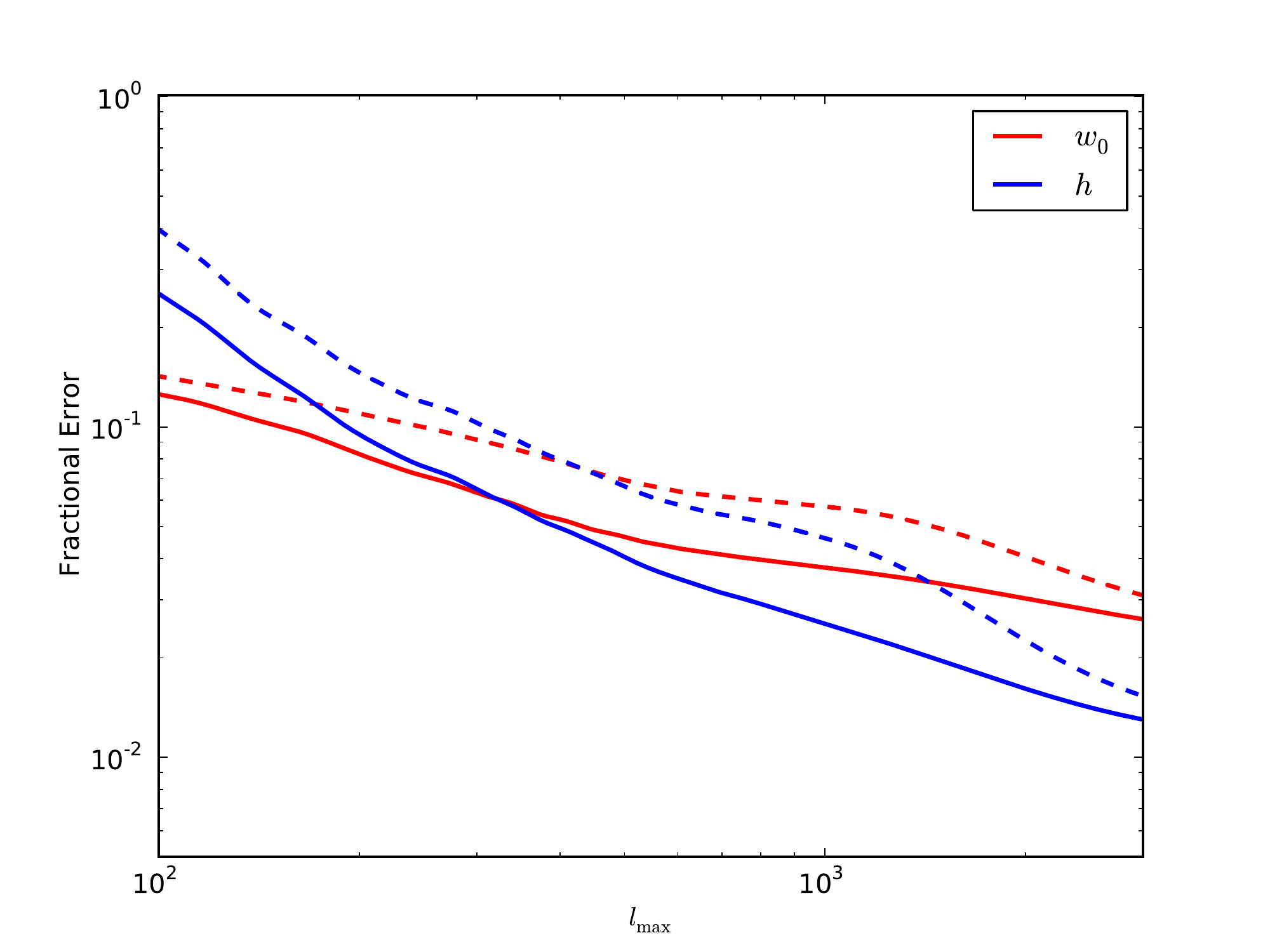}
\caption{Projected fractional constraints of two cosmological parameters (the Hubble constant $h$ and dark energy equation of state $w$) from lensing in the Dark Energy Survey (5000 square degrees) and an overlapping CMB lensing survey that covers 1000 square degrees and has noise of 2.5 $\mu$K-arcmin. All sets of curves include priors from Planck and marginalize over other cosmological and nuisance parameters. The constraints are shown as a function of the maximum multipole moment (smallest scale) that will be used in the analysis ($l_{\rm max}$ in the range of $1000-3000$ is often assumed). Dashed curves shows the constraints if the two surveys did not overlap and solid if they do overlap (in which case cross-correlation information may be utilized). Adapted from~\cite{Vallinotto:2013eva}.}
\label{fig:alberto}
\end{center}
\end{figure}

\subsection{Early Dark Energy via the CMB}
\label{sec:synergy}

Further synergies exist between data sets, e.g.\ from ground and space or
at different wavelengths, beyond formal cross-correlations.  We briefly
mention some examples, focusing on the impact on cosmological parameter
estimation.

A combination of all sky, space-based and high resolution, low noise
ground-based CMB measurements provide leverage on early dark energy, for
example.  A number of models, particularly high energy physics models with
early attractor solutions, allow for dark energy at a low level (but
millions of times higher than in quintessence theories) during the early
universe.  Such persistence can be one of the key properties for exploration
of the origin of dark energy.

Focusing on the Doran-Robbers model \cite{doranrobbers}, the dark energy
density
\be
\Omega_{de}(a)=\frac{1-\om-\Oe(1-a^{-3w_0})}{1-\om(1-a^{3w_0})}
+\Oe(1-a^{-3w_0}) \ ,
\end{equation}
containing two new parameters: the early dark energy density $\Oe$ and the
present dark energy equation of state $w_0$. At late times ($z<2$)
distances in this model can be mimicked by standard
dark energy with  a time varying equation of state, $w_a=5\Omega_e$,
so 0.5\% early dark energy acts like $w_a=0.025$,
indistinguishable by late time experiments. CMB measurements can break this degeneracy.

We can project constraints on the two dark energy parameters, together
with the standard six vanilla parameters and the sum of neutrino masses
$m_\nu$.  Fiducial values of the new parameters are taken as $\Oe=0.009$,
$w_0=-0.97$, $m_\nu=0.1\,$eV.
The projected data includes Planck and ground based CMB data.
Systematics are treated by imposing a high multipole cutoff
of $l_{\rm max}=3000$ on both temperature and polarization.

\begin{table*}[!htb]
\begin{tabular}{l|ccccccccc}
Case&$10^5\ob$ &$10^4\ocd$ &$m_\nu$(meV)&$\om$&$10^3 \tau$&$10^3 n_s$&$\sigma_8$&$w_0$&$\Oe$\\
\hline
Planck& 13.0& 15.2& 165& $\ 0.0534\ $& 4.32& 3.36& $\ 0.0552\ $& $\ 0.214\ $& $\ 0.00455\ $\\
Pl+4000& 6.75& 10.8& 110& 0.0445& 3.75& 2.56& 0.0459& 0.177& 0.00319\\
Pl+10000& 5.13& 9.22& 90& 0.0378& 3.24& 2.26& 0.0390& 0.151& 0.00258\\
Pl+15000& 4.46& 8.38& 80& 0.0341& 2.94& 2.09& 0.0352& 0.137& 0.00229\\
Pl+10k$@$1$\mu$K-arcmin& 3.77& 8.58& 80& 0.0332& 3.19& 2.17& 0.0348& 0.133& 0.00228\\
Pl+10k, $l^P_{\rm max}=5000\ $& 4.46& 8.73& 86& 0.0332& 3.23& 2.21& 0.0347& 0.133& 0.00247\\
\hline
Gain& 2.53& 1.65& 1.83& 1.41& 1.33& 1.49& 1.42& 1.42& 1.76
\end{tabular}
\caption{$1\sigma$ constraints from future CMB experiments on cosmological
parameters, including neutrino mass and early dark energy.  Gain shows the
improvement factor when adding 10000 deg$^2$ of high resolution, low noise ground
based data to Planck.
}
\label{tab:cases}
\end{table*}

Table~\ref{tab:cases} shows the projected cosmological constraints in the early
dark energy plus neutrino mass model for Planck and several future
experiments.  Planck full temperature and polarization data in the future
can achieve a 68\% CL constraint on $\Omega_e$ of 0.45\%,
when also marginalizing over the other cosmological parameters, the sum of
neutrino masses, and the present equation of state $w_0$.  Adding 10,000
square degrees of high resolution (1 arcmin), low noise (7 uK-arcmin in
polarization) ground based data improves this limit to 0.26\%, while
simultaneously constraining the sum of neutrino masses to 90 meV (rather
than 165 meV from Planck alone in this model).  Note that since both early
dark energy and neutrino mass suppress early growth, constraints on neutrino
mass tend to be tighter in non-early dark energy models; thus early dark
energy models give more conservative bounds.  Planck data could also
distinguish a low sound speed from quintessence's $c_s=1$ at better than
99\% CL \cite{1010.5612}.

\subsection{Multi-wavelength Studies of Galaxy Clusters}
\label{sec:cc-clusters}

Galaxy clusters are perhaps the quintessential example of the value of observing cosmic phenomena at different wavelengths. Recall that one of the key uses for clusters in dark energy studies is to compare the abundance above a given mass threshold as a function of redshift with theoretical predictions. The key uncertainty remains the mass determination, and it is in this regard that multi-wavelength studies become particularly important.

The three established ways to find galaxy clusters are via the X-ray emission
of their hot ($>$ keV) gas; the signature of inverse Compton scattering on the
CMB (the so-called Sunyaev-Zel'dovich (SZ) effect), and by identifying
overdensities of galaxies in optical surveys (see Figure~\ref{fig:sptcl}). The full exploitation of cluster
counts for cosmological studies requires a multi-wavelength approach.  The
primary requirements are a wide-area cluster survey with a well-understood
selection function and high-quality mass measurements. Optical observations
play a central role in the assembly of large cluster catalogs and also,
uniquely with current instrument (i.e. before WFIRST-AFTA), in the provision of redshift information and weak gravitational
lensing data, one of the key requirements for robust mass
calibration~\cite{vaaka12}.  X-ray observations also play a crucial role, enabling the
construction of cluster catalogs of exquisite purity and completeness and
providing unparalleled  low-scatter mass proxy information, which can boost the
cosmological constraining power of a survey by a factor of a few with respect
to self calibration alone~\cite{cunha09,wuetal10,mantz10}.  Millimeter
wavelength observations provide a third pillar of cluster research, enabling
efficient searches for clusters at high redshifts, beyond the reach of optical
surveys~\cite{bddra13}.  In combination, these complementary measurements
offer outstanding opportunities to advance our understanding of dark energy and
fundamental physics.


CMB experiments will enable two new useful sets of information about clusters.  First, clusters can be cross-correlated with the CMB lensing maps mentioned in \S\ref{sec:cc-cmb} to measure the linear bias of clusters as tracers of the total matter distribution.  Because the cluster bias is a monotonic function of cluster mass, this measurement also allows one to place constraints on cluster masses, thereby further empowering cluster abundance studies.  The first such detection was recently achieved using a combination of {\it Planck} and SDSS data \cite{2013arXiv1303.5080P}.  The constraint is currently limited by a combination of the depth of the {\it Planck} CMB measurements and the mismatch in the redshift sensitivity of the two data sets, with the SDSS cluster sample peaking at a relatively low redshift of $z \approx 0.2$.  Both will be significantly improved using future data sets, with current generations CMB experiments (e.g., SPTpol, ACTpol) achieving an order of magnitude deeper measurements and future optical and near infrared  cluster catalogs (e.g., DES, HSC, LSST, Euclid, WFIRST-AFTA) extending to $z \sim 1$.   Further improvements should be achieved with next generation CMB experiments.

The second type of cross-correlation analysis is to directly measure the gravitational lensing of the CMB by clusters.  This can be done with either CMB temperature or polarization data, however the best sensitivity will ultimately be achieved with polarization measurements \cite{huetal07}.  Current generation polarization experiments (e.g., SPTpol, ACTpol) are expected to constrain cluster masses with $\sim$10\% precision for every $\sim$10$^3$ galaxy clusters.  A next-generation Stage 4 CMB experiment would achieve another order of magnitude deeper CMB data, which may result in making CMB lensing competitive with future optical and near infrared experiments, depending on the systematics floor reached by LSST, WFIRST-AFTA, and Euclid.


Just as important as the reduced statistical scatter, multi-wavelength data allow for improved control of systematics, and for critical, non-trivial self-consistency
tests that can signal the presence of systematic errors.  As a simple example, consider cluster centering.  In order to derive weak lensing
masses, or for cross correlating galaxy clusters with other samples, one must specify the center of the cluster, which is difficult to do in optical,
but less so in X-rays and/or SZ.  Consequently, one can utilize X-ray/SZ data to characterize the mis-centering distribution in optical samples:
optical can given the location of a cluster along the line of sight, while X-rays/SZ are better at locating the cluster on the sky.  Likewise,
one could imagine a cosmological model, and two sets of scaling relations --- say optical--mass and SZ--mass --- which simultaneously
reproduce the abundance function in the optical, and the abundance function in SZ.  One can then string these scaling relations to
predict the optical--SZ scaling relation, and compare that to what is observed.  If there is no agreement, then the model must be ruled
out, betraying an error in either the cosmological model, or either set of scaling relations.
Indeed, these type of non-trivial constraints have already been seen to fail in current samples \cite{planck11optical}, and have been
used to argue for the presence of systematics in various data sets \cite{rozoetal12d}.   As cosmological constraints become increasingly
refined, these types of systematics cross checks are bound to become increasingly important in establishing the robustness of cosmological
constraints from galaxy clusters.

\begin{figure}[thbp]
\begin{center}
\includegraphics[width=.6\columnwidth]{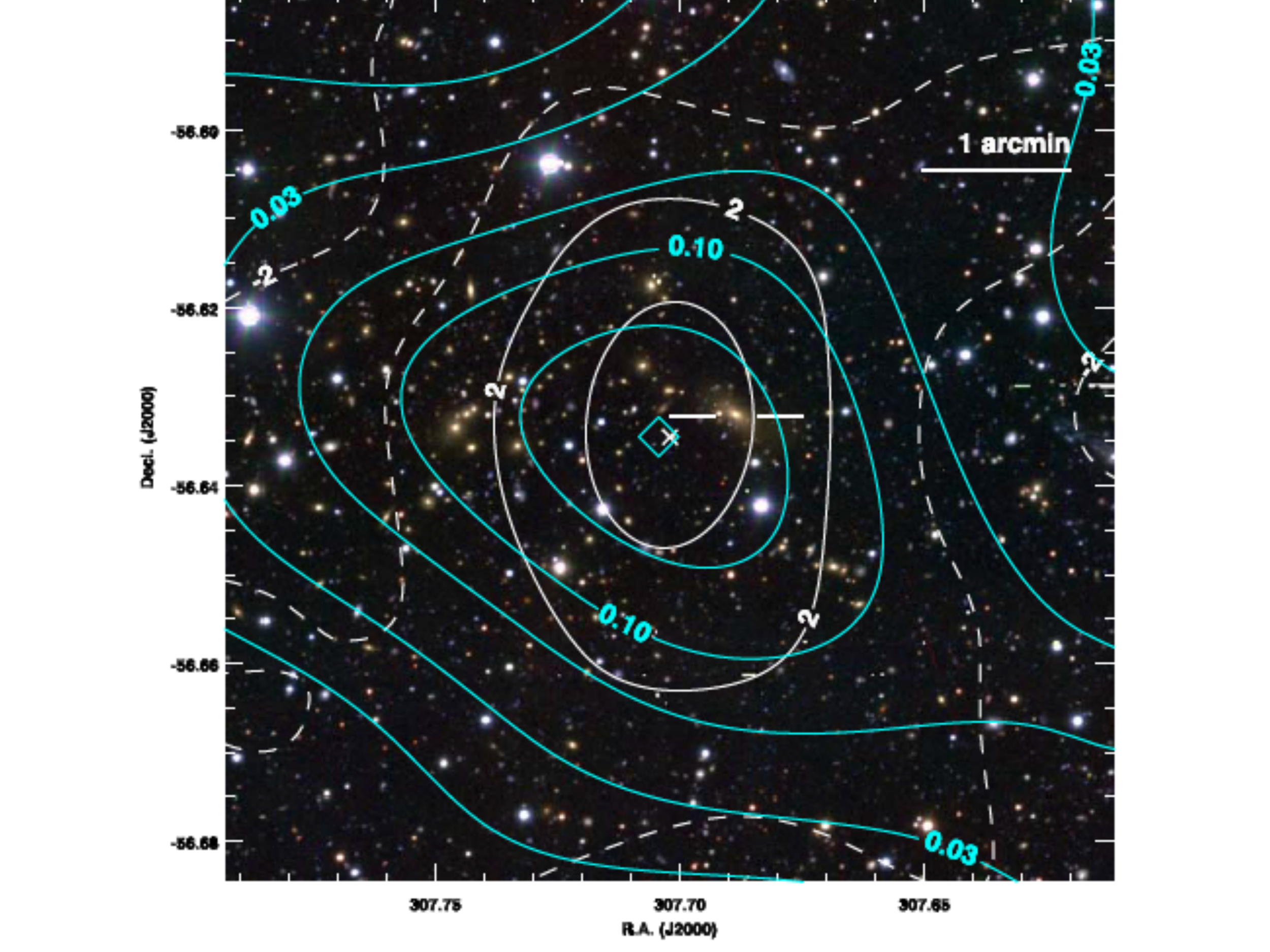}
\caption{Map of a galaxy cluster~\cite{High:2012un} using three probes: (i) weak gravitational lensing (blue contours with labels showing the projected density $\kappa$); (ii) hot gas as measured by the Sunyaev-Zel'dovich distortion of the CMB (white contours with labels giving signal to noise); and (iii) galaxies as observed in three optical bands (background).  Each of these measurements provides complementary information about galaxy clusters being studied; combining them is key for obtaining robust cosmological results from cluster counts.}
\label{fig:sptcl}
\end{center}
\end{figure}

By using X-ray and optical observations of galaxy clusters in concert, we can
also measure the distance of clusters directly, independent of redshift
allowing the acceleration of the Universe to be traced~\cite{2011ARA&A..49..409A, ammak13}. The
matter content of the most massive clusters provides a fair sample of the
matter content of the Universe.  The ratio of baryonic-to-total mass should
match the ratio of cosmological parameters $\Omega_b /\om$.  As the baryonic mass
content of clusters is dominated by the X-ray gas, the X-ray gas fraction,
$f_{\mathrm{gas}}$, is an effective cosmological probe.  Simulations show that the $f_{\mathrm{gas}}$
parameter should be approximately constant with redshift, yet $f_{\mathrm{gas}}$ also
depends on the assumed distance to the clusters.  Therefore, measurements of
$f_{\mathrm{gas}}$ allow the distances of clusters to be inferred, as
illustrated in Figure~\ref{fig:fgas}.

\begin{figure}[thbp]
\begin{center}
\includegraphics[width=0.8\columnwidth]{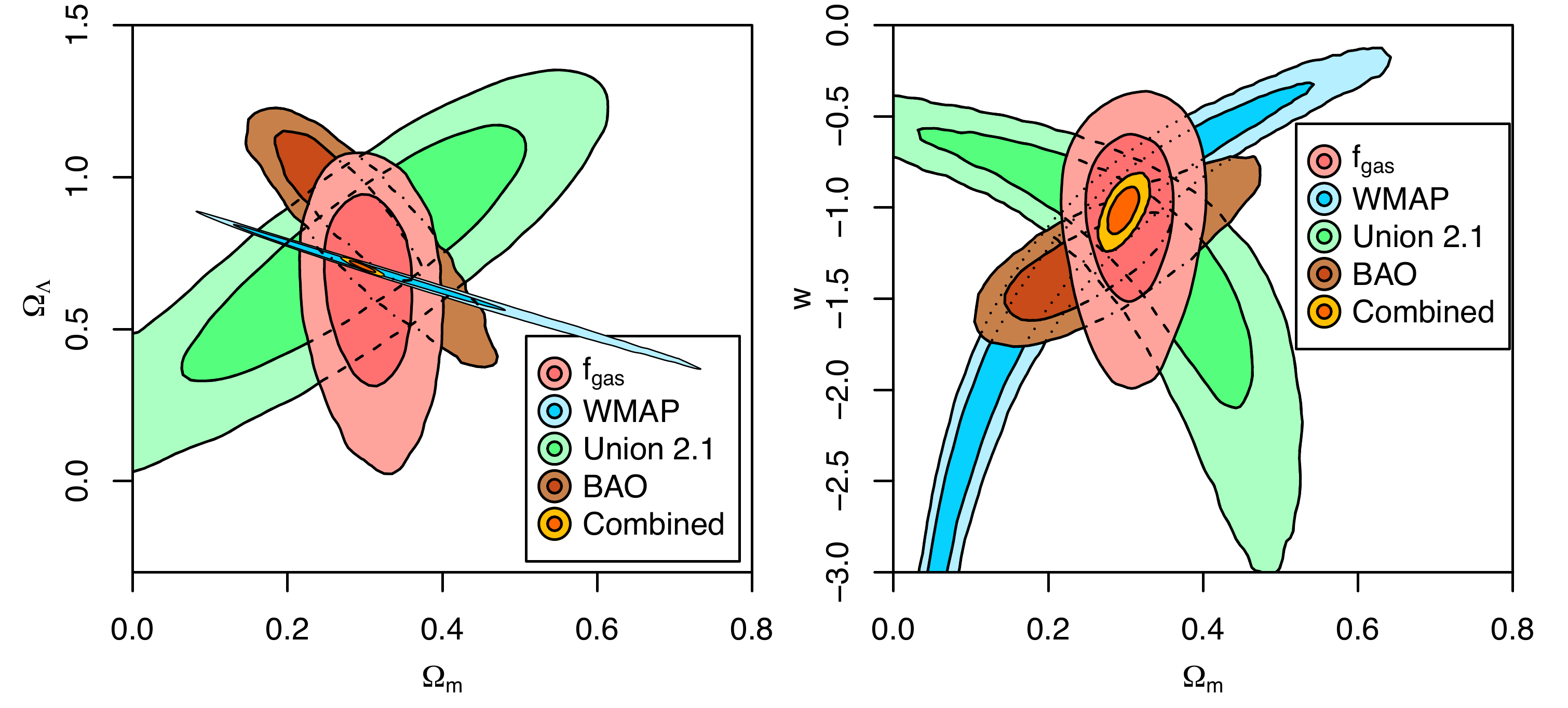}
\caption{Recent un-blinded, initial cosmology results from cluster distance
  measurements (Mantz et al., in preparation). Panels show the joint 68.3\% and
  95.4\% confidence constraints for (Left) non-flat ΛCDM and (Right) flat w-CDM
  (right) models.  Red shading shows the confidence regions obtained with
  cluster data (fgas). Independent constraints from CMB
  measurements~(blue; from \cite{hlksb12}), SN Ia~(green; from \cite{srlaa12}), and
  BAO~(brown; from \cite{aabbb13}) are also shown. Note the tight constraints on
  $\Omega_m$ from clusters, which are largely independent of the cosmological model
  assumed.}
    \label{fig:fgas}
  \end{center}
  \end{figure}

Unlike growth measurements, which must be carried out with complete cluster
samples, with distance measurements one is free to 'cherry-pick' the best
targets (the most luminous and dynamically relaxed clusters).  Recent progress
has been made in the automated selection of relaxed clusters, to maximize the
information to be extracted from existing data archives.  Secondly, detailed
hydrodynamic simulations have been utilized to help determine the optimal
measurement radii to maximize the cosmological signal and minimize systematic
uncertainties.  A third key step has been the incorporation of rigorous weak
lensing mass calibration in the analysis~\cite{avkaa12}.  In this way, a
multi-wavelength approach directly addresses the main systematic uncertainty in
the experiment, namely the hydrostatic mass bias arising from non-thermal
pressure support of the X-ray emitting gas.

A second way of using clusters to measure cosmic distances utilizes combined
X-ray and mm-wavelength measurements of cluster pressure
profiles~\cite{sw78}. Given an observed SZ signal at mm-wavelengths, combined
with the predicted SZ signal from the X-ray measurements, given the distance
dependence of the X-ray measurements, one can solve for $d_A$.  Fortunately, the
ideal clusters for this test are the same ones needed for the
$f_{\mathrm{gas}}$ test.  Although this "XSZ" test is less sensitive than the
$f_{\mathrm{gas}}$ experiment, progress is being made~\cite{ammak13}.
In particular, upcoming X-ray surveys such as eROSITA~\cite{mpbbb12}
will discover thousands of hot, massive clusters out to high redshift, and the
best candidates can be followed up with deep targeted X-ray and SZ
observations.


\section{Enabling Joint Analyses}
\label{sec:cc-joint}
The projects running now and being planned for the coming decade will produce a rich legacy data set toward the end of the 2020s.  Given the current theoretical uncertainly about the true nature of dark energy, and the multiple systematics that plague dark energy measurements, the joint analysis of data sets across a wide range of wavelengths, observatories, and techniques will certainly be required to achieve the tightest possible cosmological constraints. There will be inevitable synergies in data sets.  For example:
\begin{enumerate}
   \item Spectroscopic redshifts from DESI, Euclid, PFS, and WFIRST-AFTA as well as  thirty meter class telescopes will be used to calibrate photometric redshifts for  LSST, Euclid, and WFIRST-AFTA..
   \item The best possible photometric redshifts will come from combining deep ground based photometry in the optical from LSST and near-infrared data from space missions including WFIRST-AFTA and Euclid.
   \item The calibration of weak lensing shapes will benefit from cross-correlating deep ground based images (LSST) and higher resolution images taken with the stable point spread function only available in space (Euclid and WFIRST-AFTA).
   \item Many systematic effects can best be mitigated by comparing multiple measurements.  For example, imaging will be performed with some combination of CCDs and near infrared detectors.  Each type of detector has different non-idealities that affect photometry and galaxy shape measurement. Cross comparison of Euclid, LSST, and WFIRST-AFTA galaxy shapes and photometry will provide the highest precision measurements of those quantities.
\end{enumerate}

At each stage, the tightest constraints on dark energy  come from careful \emph{joint} analyses of all extant, relevant data sets.  Joint analyses that use ``catalog level'' data from each experiment are relatively straightforward, but the data reduction and analysis procedures that are optimized for the internal use of a single data set may not provide the optimal images, spectra and catalogs for such a joint analysis. For some purposes therefore, it may be necessary to process images and spectra  from raw data in a way that optimizes the accuracy of shape and redshift information for each galaxy.   For example, when calculating photometric redshifts using a combination of ground and space-based data, apertures might be used that would not be the optimal apertures for either data set alone.  Optimal joint analysis requires careful thought and preparation.  This will also require cooperation among different consortia and government agencies.  We urge the consortia and agencies to begin planning for these activities now, since building these capabilities into the various pipelines from the start may be cheaper and more efficient then adding them after the fact.  We propose a few ideas that would enable the best possible use of the rich and varied data sets enabled by the projects discussed here:
\begin{enumerate}
   \item All data sets should eventually be available to all members of the international astronomy and physics community.  Joint analyses will likely be an international effort.  While many experiments will necessarily have a proprietary period for their data, these periods should be limited and every effort should be made to serve the data to the public after the proprietary period.
   \item Care must be taken to make sure the data being served to the public is served in a useful fashion.  There is a large difference between data that is available and data that is useful.  Proper tools must be provided for the access, querying, and downloading of calibrated images and  spectra as well as properly documented catalogs and derived data products.  Consortia and agencies should make sure to budget for the user support needed to enable the joint analyses discussed here.
       \item Given the large volume of data that will be produced by future missions and surveys, careful thought must be given about how to disseminate images and catalogs. End users across the world will need to access and analyze the data.  Care must be taken to ensure proper bandwidth.  Likewise, data products should be optimized to minimize what needs to be transferred.  This will require carefully planning as well as diligence in describing the provenance of data products.
   \item The consortia that build, operate, and analyze the data from each survey, mission, or telescope will be the experts on that particular data set.  Agencies should begin to think about how to fund and organize the analysis of data products from multiple experiments and countries that preserves the knowledge of the original consortia and incorporates that knowledge into subsequent data reduction and analyses.
\end{enumerate}
Given that some of the relevant experiments are already underway, the time is ripe for consortia and agencies to consider the optimal steps to ensure access to disparate data sets and the usefulness of data products for complete but necessary joint reduction and analyses. The SDSS database, the Mikulski Archive for Space Telescopes (MAST), and the Infrared Science Archive (IRSA) provide valuable experience to draw from, demonstrating both the large effort required to produce a powerful and well documented database and the enormous benefit that comes from doing so.

\section{Conclusions}
\label{sec:cc-conclusions}

In this white paper, we have presented a wide variety of ways in which information from multiple probes of dark energy may be combined to obtain additional information not accessible when they are considered separately.  Methods based upon cross-correlation statistics have great power to break degeneracies and test for systematics; however, their potential is only beginning to be realized.  By bringing in information from other wavelengths, the capabilities of the existing probes of dark energy can be enhanced and systematic effects can be mitigated further.  Given the scope of future dark energy experiments, the greatest gains may only be realized with more coordination and cooperation between multiple project teams; we recommend that this interchange should begin sooner, rather than later, to maximize scientific gains.


{\it Acknowledgements.} The authors gratefully acknowledge helpful contributions from and productive discussions
with Matthew Becker, Tomasz Biesziadzinski, Brandon Erickson, August
Evrard, Jaime Forero-Romero, Dragan Huterer, Stephen Kent, James Kowalkowski, Rachel Mandelbaum, Michael
Mortonson, Marc Paterno, Adrian Pope, Erin Sheldon and Matthew Turk. JR is supported by JPL, run under a contract for NASA by Caltech.

\bibliography{cc}{}




\end{document}